\documentclass[
aps,
prc,
superscriptaddress,
floatfix,
nofootinbib,
twocolumn
]{revtex4-2}

\usepackage{amsmath,amssymb,amsfonts,physics,graphicx,float}
\graphicspath{{pic/}}
\usepackage[setpagesize=false]{hyperref}
\allowdisplaybreaks[4]

\usepackage[labelformat=simple]{subcaption} 

\usepackage{ulem}
\usepackage{color}
\definecolor{ar}{rgb}{1.0, 0.01, 0.24}
\definecolor{al}{rgb}{0.82, 0.1, 0.26}
\definecolor{ev}{rgb}{0.56, 0.0, 1.0}

\newcommand{\beq}{\begin{eqnarray}}
\newcommand{\eeq}{\end{eqnarray}}

\newcommand{\lag}{\mathcal{L}}

\newcommand{\Interp}{\mathrm{I}}

\newcommand{\rmd}{ \mathrm{d} }
\newcommand{\vx}{ {\bf x} }
\newcommand{\vp}{ {\bf p} }

\begin{document}


\title{ 
Chiral condensates for neutron stars in hadron-quark crossover: \\
from a parity doublet nucleon model to an NJL quark model
}

\author{Takuya Minamikawa}
\email{minamikawa@hken.phys.nagoya-u.ac.jp}
\affiliation{Department of Physics, Nagoya University, Nagoya 464-8602, Japan}

\author{Toru Kojo}
\email{torujj@mail.ccnu.edu.cn}
\affiliation{Key Laboratory of Quark and Lepton Physics (MOE) and Institute of Particle Physics,
Central China Normal University, Wuhan 430079, China}

\author{Masayasu Harada}
\email{harada@hken.phys.nagoya-u.ac.jp}
\affiliation{Department of Physics, Nagoya University, Nagoya 464-8602, Japan}
\affiliation{Kobayashi-Maskawa Institute for the Origin of Particles and the Universe, Nagoya University, Nagoya 464-8602, Japan} 
\affiliation{Advanced Science Research Center, Japan Atomic Energy Agency, Tokai 319-1195, Japan}

\date{\today}

\begin{abstract}
We study the chiral condensates in neutron star matter from nuclear to quark matter domain. 
We describe nuclear matter with a parity doublet model (PDM), quark matter with the Nambu--Jona-Lasino (NJL) model, 
and a matter at the intermediate density by interpolating nuclear and quark matter equations of state. 
The model parameters are constrained by nuclear physics and neutron star observations.
Various condensates in the interpolated domain are estimated from the chemical potential dependence of the condensates at the boundaries of the interpolation.
The use of the PDM with substantial chiral invariant mass ($m_0 \gtrsim 500$ MeV, 
which is favored by the neutron star observations) predicts the mild chiral restoration, and the significant chiral condensate remains to baryon density $n_B \sim 2-3n_0$ ($n_0\simeq 0.16\,{\rm fm}^{-3}$: 
nuclear saturation density), smoothly approaching the NJL predictions for the color-flavor-locked phase at $n_B \gtrsim 5n_0$. 
The same method is applied to estimate diquark condensates, number densities of up-, down- and strange-quarks, and the lepton fraction. 
In our descriptions the chiral restoration in the interpolated domain proceeds with two conceptually distinct chiral restoration effects;
 the first is associated with the positive scalar density in a nucleon, 
 relevant in dilute regime, and the other primarily arises from the modification of the quark Dirac sea, which is triggered by the growth of the quark Fermi sea.
We discuss several qualitative conjectures to interpolate the microphysics in nuclear and quark matter.
\end{abstract}

\maketitle


\section{Introduction}

The chiral condensates are important quantities to characterize the state of matter in Quantum Chromodynamics (QCD).
While its high temperature behavior at low baryon density has been studied 
in the lattice Monte Carlo simulations \cite{Aoki:2009sc,Bonati:2015bha,HotQCD:2018pds} 
its behavior at low temperature and high baryon density remains unknown due to the lack of reliable first principle computations.

In this situation recent neutron star (NS) observations, which tightly constrain the equations of state (EOS) of matter, 
are useful tools to access the properties of dense matter (see, e.g., Ref.\cite{Baym:2017whm} for a review and Ref.\cite{Kojo:2020krb} for a mini-review). 
The existence of two-solar mass ($2M_\odot$) NSs with
$1.91  M_\odot$ \cite{Arzoumanian:2017puf,Fonseca:2016tux,Demorest:2010bx},
$2.01 M_\odot$ \cite{Antoniadis:2013pzd}, and
 $M=2.08\pm 0.07 M_\odot$  \cite{Fonseca:2021wxt}, 
 the radius constraint $R_{1.4} \lesssim 13$ km for $1.4M_\odot$ NS deduced from the gravitational event GW170817 \cite{TheLIGOScientific:2017qsa}, 
 and recent NICER constraints on $R_{1.4} \simeq R_{2.08} \simeq 12-13$ km \cite{Miller:2021qha,Riley:2021pdl,Raaijmakers:2021uju}, 
disfavor the strong first order phase transitions for the domain between the nuclear saturation density ($n_0 \simeq 0.16\,{\rm fm}^{-3}$) and the core density achieved in the $2M_\odot$ NSs. 
The core baryon density of $2M_\odot$ NS is $n_B \gtrsim 4-5n_0$, presumably high enough to apply quark matter descriptions, so we infer that nuclear matter smoothly transforms into quark matter (modulo weak first order phase transitions).

Following the three-window picture developed in Refs.\cite{Masuda:2012kf,Masuda:2012ed,Masuda:2015kha},
we have been constructing series of unified EOS covering from nuclear to quark matter, and used the NS observations to constrain them \cite{Kojo:2014rca,Fukushima:2015bda,Baym:2019iky,Minamikawa:2020jfj}. 
We use a nuclear model at $n_B \lesssim 2n_0$, a quark model at $n_B\gtrsim 5n_0$, and then interpolate them to construct EOS for $n_B \simeq 2-5n_0$.
The resulting EOS can be made consistent with available NS constraints by choosing the proper range of parameters in our model.
Meanwhile, the real strength of our approach is that, 
with the microscopic degrees of freedom being manifest, we can use the model to predict physical quantities other than EOS and get some insights on the microphysics \cite{Baym:2017whm}. 

The previous works based on interpolation 
schemes \cite{Masuda:2012kf,Masuda:2012ed,Masuda:2015kha,Kojo:2014rca,Fukushima:2015bda,Baym:2019iky,Minamikawa:2020jfj}, 
however, did not manifestly address the microscopic quantities in the interpolated domain, and thus did not utilize the full potential of the framework.
In this paper, we examine various condensates and matter compositions by constructing {\it unified generating functionals} for the pressure, including external fields coupled to condensates.
The condensates are obtained by differentiating the functionals with respect to the external fields. 
In this way it is possible to examine the chiral and diquark condensates as well as lepton fractions from nuclear to quark matter domain.
The condensates in the interpolated domain are affected by the physics of nuclear and quark matter through the boundary conditions for the interpolation.
We examine condensates in nuclear and quark matter models in detail, and then infer the physics relevant in the interpolated domain.

In this work, a nuclear matter is described by a parity double model (PDM) for nucleons \cite{Detar:1988kn,Jido:2001nt} coupled to the meson potentials. 
(For nuclear matter descriptions based on the PDM, see e.g. 
Refs.\cite{Hatsuda:1988mv,Zschiesche:2006zj,Dexheimer:2007tn,Dexheimer:2008cv,Sasaki:2010bp,Sasaki:2011ff,Gallas:2011qp,Paeng:2011hy,Steinheimer:2011ea,Dexheimer:2012eu,Paeng:2013xya,Heinz:2013hza,Motohiro:2015taa,Benic:2015pia,Mukherjee:2016nhb,Mukherjee:2017jzi,Suenaga:2017wbb,%
Takeda:2017mrm,Marczenko:2017huu,Paeng:2017qvp,Marczenko:2018jui,Abuki:2018ijb,Takeda:2018ldi,Yamazaki:2019tuo,Harada:2019oaq,Marczenko:2019trv,Harada:2020etl,Marczenko:2020jma}).
In the PDM, nucleons with positive and negative parities form a doublet in which the masses are split by the chiral condensates; 
in chiral symmetric phase the masses get degenerated with the finite chiral invariant mass, $m_0$. 
Finite temperature lattice simulations indicated the parity degeneracy of nucleons at finite mass \cite{Aarts:2017rrl}, supporting the idea of the chiral invariant mass.
With a larger $m_0$, the nucleon mass can reach the experimental value within weak couplings between nucleons and the chiral condensates;  
in turn the chiral condensates become less sensitive to changes in the nuclear medium, 
and hence the chiral restoration driven by the increase of density proceeds more mildly.
This feature also affects the structure of neutron stars.
In the context of the $\sigma$-$\omega$-$\rho$ type models \cite{Walecka:1974qa,Serot:1984ey,Serot:1997xg} for neutron star matter, 
the reduced $\sigma N$ coupling leads to a less attractive $\sigma$ exchange, 
so that a weaker repulsion by the $\omega$ exchange is sufficient to reproduce the physics near the saturation density. 
This trends becomes more important at higher baryon density where the $\sigma$ fields decrease but $\omega$ fields increase. 
In the context of NS observations, 
the reduction of $\omega$-repulsion reduces NS radii from $13-14$ km to $12-13$ km \cite{Minamikawa:2020jfj}.

A quark matter is described by the standard Nambu--Jona-Lasinio (NJL) model \cite{Hatsuda:1994pi} plus additional effective interactions, the vector repulsion and diquark attraction \cite{Kojo:2014rca}. 
The vector repulsion is particularly important in the context of the chiral restoration, as it  tempers the growth in density, smoothing out the chiral restoration \cite{Kitazawa:2002jop,Bratovic:2012qs}.
Meanwhile the inclusion of diquark terms leads to the color-flavor-locked (CFL) phase in which $ud$-, $ds$-, and $su$-diquark pairs condense, 
and these pairings favor a larger quark Fermi sea \cite{Alford:2007xm}. 
After including these two competing effects, 
we found that the substantial amount of the chiral condensates remain in quark matter, 
and the effective quark masses at $n_B \sim 5n_0$ for up- and down-quarks are $\sim 50$ MeV, for strange quarks $\sim 300$ MeV \cite{Baym:2017whm}. 

The chiral condensates in the interpolated domain, which are influenced by matching to nuclear and quark matter models,
are supposed to contain the mixture of two distinct chiral restoration effects. 
The first type is the chiral restoration due to the positive scalar charge in a nucleon; 
the sign is opposite to the negative vacuum chiral density so that these charges cancel in spatial average, leading to the reduction of chiral condensates.
This sort of chiral restoration does not necessarily demands the knowledges about the structural changes of hadrons in medium, 
and the estimate is valid at least in dilute regime.
Meanwhile, the NJL quark models describe more fundamental changes associated with changes in the Dirac sea or in the constituent quark properties \cite{Baym:2017whm}. 
Implementing such information through the high density boundary conditions introduces a qualitative trend that differs from the high density extrapolation of the PDM.

The recipe to calculate chiral condensates in the interpolated domain is also usable for other condensates, such as diquark condensates and quark densities for various flavors.
In particular the strangeness and lepton fractions have distinct trends from those in purely hadronic models.

This paper is structured as follows.
In Sec.\ref{sec:nuclear_matter}, we discuss the PDM as a nuclear matter model. The equations of state and chiral condensates for light ($u,d$) and strange quarks are discussed.
In Sec.\ref{sec:quark_matter}, we discuss the NJL model with effective diquark and vector interactions.
In Sec.\ref{sec:unified_EOS}, we discuss the recipe to compute condensates in the interpolated domain, and examine the behaviors of various condensates.
In Sec.\ref{sec:discussions}, we give several conjectures to describe our results in the microphysics language.
Sec.\ref{sec:summary} is devoted to the summary. 

In this paper we write the integrals over space, $\int_{x} = \int \rmd^4 x$, $\int_{\vx} = \int \rmd^3 \vx$, and over momentum, $\int_{\vp} = \int \frac{\, \rmd^3 \vp \,}{\, (2\pi)^3 \,}$.


\section{Nuclear matter} \label{sec:nuclear_matter}

\subsection{Parity Doublet Model (PDM)} \label{sec:PDM}

The mean field grand potential functional in our PDM consists of the meson and baryon contributions \cite{Minamikawa:2020jfj},
\begin{eqnarray}
\Omega_{\rm bare} (\mu_B,\mu_Q ;\varphi )
= \Omega_M ( \varphi  ) +\Omega_B (\mu_B,\mu_Q; \varphi ) \,,
\end{eqnarray} 
where $\varphi$ collectively denotes the meson fields $(\sigma, \omega, \rho)$,
and $\mu_B$ and $\mu_Q$ are 
the baryon number and the electrical charge chemical potential, respectively.
The meson part is
\begin{eqnarray}
 \Omega_M
=
&-& \frac12\bar\mu^2\sigma^2
+ \frac14\lambda_4\sigma^4
- \frac16\lambda_6\sigma^6
- m_\pi^2 f_\pi \sigma 
\nonumber \\
& - & \frac12m_\omega^2\omega^2
-\frac12m_\rho^2\rho^2 \,,
\end{eqnarray}
where $m_\pi$ and $f_\pi$ are the pion mass and decay constant in vacuum, respectively, 
and the baryon part is ($\Theta(x)$ is the step function)
\begin{eqnarray}
\Omega_B 
=-2 \sum_{i=\pm}\sum_{\alpha=p,n }
\int_\mathbf{p}(\mu^\ast_\alpha-E_\mathbf{p}^i)  \Theta(\mu^\ast_\alpha-E_\mathbf{p}^i)\ \,,
\label{eq:omega_B}
\end{eqnarray}
where the index $i=\pm$ is for the parities of nucleons and $\alpha$ for protons and neutrons.
Here we used the no-sea approximation which neglects the Dirac sea contributions from nucleons.
The single particle energies for nucleons 
and chemical potentials are respectively given by 
\begin{eqnarray}
&&E_\mathbf{p}^i=\sqrt{\mathbf{p}^2+m_i^2} \,, \\
&&\mu^\ast_p
=\mu_B - g_{\omega NN}\omega - \frac{1}{\, 2 \,} g_{\rho NN}\rho + \mu_Q \,,
\nonumber \\
&&\mu^\ast_n
= \mu_B - g_{\omega NN}\omega +\frac{1}{\, 2 \,} g_{\rho NN}\rho \,.
\end{eqnarray}
In the PDM, there are doublet fields $N_1$ and $N_2$ that couple to the $\sigma$ 
with the couplings $g_1$ and $g_2$. 
By calculating the eigenstates for coupled equations of $N_1$ and $N_2$, the masses for positive and negative parity nucleons are given by the following form,
\begin{eqnarray}
m_\pm \equiv \sqrt{m_0^2+\qty(\frac{g_1+g_2}{2})^2\sigma^2} \mp \frac{|g_1-g_2|}{2}\sigma \,,
\end{eqnarray}
where $m_0$ is the chiral invariant mass. 
Actually $m_-$ does not appear for $n_B \le 2n_0$ in our choices of $m_0$, so we will use the notation $m_N$ for the positive parity nucleon mass $m_+$; $m_N \equiv m_+$.
The parameters used in our analyses and the output are summarized in Tables.\ref{input: mass}, \ref{output}, and \ref{input: saturation}.

We minimize the functional $\Omega_{\rm bare}$ at given $\mu_B$ and $\mu_Q$ by varying fields $\varphi$, and write the solutions $\varphi_* (\mu_B, \mu_Q)$. 
The substituting $\varphi_*$ into the functional, and normalize it by subtracting the vacuum contributions, our thermodynamic potential for the hadronic part is given by
\begin{eqnarray}
\Omega_h (\mu_B, \mu_Q) 
\equiv \Omega_{\rm bare} (\mu_B, \mu_Q; \varphi_*) -  \Omega_{\rm bare} (0,0;\varphi_{\rm vac} ) \,,
\end{eqnarray}
where we have chosen the model parameters in such a way that $\varphi_{\rm vac} = (\sigma_*, \omega_*, \rho_*)\big|_{\mu_B = \mu_Q = 0} = (f_\pi, 0,0)$.
In the following we will drop off the subscript $*$ to simplify the notation, unless otherwise stated.

Finally we impose the charge neutrality and $\beta$-equilibrium condition. The lepton part is 
\beq
	\Omega_{\rm lepton} &=& -2 \sum_{l=e,\mu} \int_{\mathbf{p}} (\mu_l-E_\mathbf{p}^l) \Theta(\mu_l - E_\mathbf{p}^l)\,,
\label{eq:lepton_EOS}
\eeq
where $\mu_l=-\mu_Q$, common for electrons and muons,
and $E_\mathbf{p}^l = \sqrt{ m_l^2 + \mathbf{p}^2 } $ with $m_e = 0.5$ MeV and $m_\mu=105.7$ MeV. 
For a given $\mu_B$, we tune $\mu_Q$ to $\mu_Q^*$  that satisfies the neutrality condition, 
\beq
\frac{\, \partial \big( \Omega_h + \Omega_{\rm lepton} \big) \,}{\, \partial \mu_Q \,} \bigg|_{\mu_Q^* }= 0 \,.
\eeq
In the following, whenever we mention EOS at a given $\mu_B$, we will implicitly choose $\mu_Q=\mu_Q^*(\mu_B)$, so that EOS depends on $\mu_B$ only.

\begin{table}\centering
	\caption{  {\small Physical inputs in vacuum in unit of MeV. }  }\label{input: mass}
	\begin{tabular}{cccccc}
		\hline\hline
		$m_\pi$ ~&~ $f_\pi$ ~&~ $m_\omega$ ~&~ $m_\rho$ ~&~ $m_+ (=m_N)$ ~&~ $m_-$\\
		\hline
		140 ~&~ 92.4 ~&~ 783 ~&~ 776 ~&~ 939 ~&~ 1535\\
		\hline\hline
	\end{tabular}
\end{table}
\begin{table}\centering
	\caption{ 
	{\small 
	Values of model parameters determined for several choices of $m_0$. 
	The values of the slope parameter $L_0$ is also shown as output. 
	}}\label{output}
	\begin{tabular}{c|ccccc}
		\hline\hline
		~ $m_0$ [MeV] ~&~ 500 ~&~ 600 ~&~ 700 ~&~ 800 ~&~ 900 \\
		\hline
		$g_1$                       &~ 9.02 ~&~ 8.48 ~&~ 7.81 ~&~ 6.99 ~&~ 5.96 \\
		$g_2$                       &~15.5  ~&~ 14.9 ~&~ 14.3 ~&~ 13.4 ~&~ 12.4\\
		$\bar\mu^2/f_\pi^2$  &~ 22.7 ~&~ 22.4 ~&~ 19.3 ~&~ 11.9 ~&~ 1.50\\
		$\lambda_4$             &~ 41.9 ~&~ 40.4 ~&~ 35.5 ~&~ 23.1 ~&~ 4.43\\
		$\lambda_6f_\pi^2$  &~ 16.9 ~&~ 15.8 ~&~ 13.9 ~&~ 8.89 ~&~ 0.636\\
		$g_{\omega NN}$     &~ 11.3 ~&~ 9.13 ~&~ 7.30 ~&~ 5.66 ~&~ 3.52\\
		$g_{\rho NN}$           &~ 7.31 ~&~ 7.86 ~&~ 8.13 ~&~ 8.30 ~&~ 8.43\\
		\hline
		$L_0$ [MeV] ~&~ 93.76 ~&~ 86.24 ~&~ 83.04 ~&~ 81.33 ~&~ 80.08\\
		\hline\hline
	\end{tabular}
\end{table}
\begin{table}\centering
	\caption{  {\small Saturation properties used to determine the model parameters: the saturation density $n_0$, the binding energy $B_0$, the incompressibility $K_0$ and the symmetry energy $S_0$.}  }
	\label{input: saturation}
	\begin{tabular}{cccc}\hline\hline
	~$n_0$ [fm$^{-3}$] ~&~ $B_0$ [MeV] ~&~ $K_0$ [MeV] ~&~ $S_0$ [MeV] ~\\
	\hline
	0.16 & 16 & 240 & 31\\
	\hline\hline
	\end{tabular}
\end{table}

\subsection{Chiral condensate in the PDM}

To calculate the chiral condensate in the PDM, we differentiate the thermodynamic potential with respect to the current quark mass. 
In our model the explicit chiral symmetry breaking enters only through the term $-m_\pi^2 f_\pi \sigma$;
we neglect the mass dependence in the other coupling constants in front of higher powers in meson fields. 
Such couplings are supposed to appear in the form of $\sim (\varphi/M_q)^n$ with the effective quark mass $M_q\simeq 300$ MeV from the integration of quarks.
Hence we expect that those couplings
depend mainly on the distance scale shorter than the physics of pions, and hence the current quark mass is expected to be a small perturbation in powers of $\sim (m_q/M_q)^n$.
Using the Gell-Mann--Oakes--Renner relation, the explicit symmetry breaking term can be written as
\begin{eqnarray}
\Omega_{\rm ESB} = - m_\pi^2 f_\pi \sigma = m_q \langle ( \bar{u} u + \bar{d} d )  \rangle_0 \frac{\, \sigma  \,}{\, f_\pi \,} \,,
\end{eqnarray}
where $m_q = (m_u+m_d)/2$ is the average of current quark masses of up and down quarks 
and $ \langle (\bar{u}u +\bar{d} d ) \rangle_0$ is the chiral condensate in vacuum. Accordingly
\beq
\langle ( \bar{u} u + \bar{d} d ) \rangle 
\equiv \pdv{\Omega_{\rm ESB}}{m_q}
=  \langle ( \bar{u} u + \bar{d} d ) \rangle_0 \frac{\, \sigma  \,}{\, f_\pi \,} \,,
\eeq
is the in-medium chiral condensate in our PDM 
(here we neglected $m_q$ dependence of $\langle ( \bar{u} u + \bar{d} d ) \rangle_0$ which is higher orders in $m_q/M_q$). 
Below we focus on how $\sigma$ changes as baryon density increases.

\subsection{Chiral scalar density in a nucleon}

For the estimate of in-medium chiral condensates, it is useful to use the scalar charge in a nucleon, $N_\sigma $. It can be related to the nucleon mass in vacuum as
\beq
N_\sigma 
&=&
\int_{  \vx }
\langle N| ( \bar{u} u + \bar{d} d ) (x) | N \rangle 
\nonumber \\
&=& \langle N | \frac{\, \partial  H_{\rm QCD} \,}{\, \partial m_q \,}  |N \rangle 
= \frac{\, \partial m_N^{\rm vac} \,}{\, \partial m_q \,}  \,,
\eeq
where $H_{\rm QCD}$ is the QCD hamiltonian, and in the last step we used the Hellmann-Feynman theorem \cite{Gubler:2018ctz}. 

In the PDM,  the nucleon mass depends on the current quark mass only through the modification of $\sigma$, so we calculate the scalar charge at vacuum as
\beq
N_\sigma
 = \frac{\, \partial m^{\rm vac}_N \,}{\, \partial m_q \,} 
 = \frac{\, \partial \sigma^{\rm vac} \,}{\, \partial m_q \,} \bigg( \frac{\, \partial m_N \,}{\, \partial \sigma \,} \bigg)_{\sigma = \sigma_{\rm vac}} \,.
 \label{eq:Nsigma}
\eeq
 When we calculate the variation of $\sigma^{\rm vac}$, we prepare the thermodynamic potentials at various $m_q$, calculate $\sigma^{\rm vac}$ for each potential, and evaluate the impact of the mass variation on $\sigma^{\rm vac}$.  
The mass derivative of $\sigma^{\rm vac}$ is proportional to the static correlator\footnote{
The mass derivative of the quark condensate is related to the (connected) scalar correlator at zero momentum,
\beq
\frac{\, \partial \langle \bar{q} q (x) \rangle \,}{\, \partial m_q \,}
&\sim& 
\int {\cal D} q {\cal D} \bar{q}
~ [ \bar{q} q (x) ] \frac{ \partial }{\, \partial m_q \,} \bigg( {\rm e}^{ - \int_{x'} m_q \bar{q} q (x') + ...} /Z \bigg)
\nonumber \\
&\sim& 
\int_{x'} \big\langle [ \bar{q} q (x) ] [ \bar{q} q (x') ] \big\rangle_{\rm conn.}
\sim
 \lim_{q\rightarrow 0} \frac{1}{\, q^2 + m_\sigma^2 \,} \,.
\eeq
} 
$\sim \langle (\sigma -\langle \sigma \rangle )^2 \rangle \sim m_\sigma^{-2}$, 
so is bigger for a smaller scalar meson mass.  

Multiplying $m_q$ to the scalar charge, we get the so-called nucleon sigma term,
\beq
\Sigma_N = m_q N_\sigma = \int_{\vec{x}} \langle N| m_q (  \bar{u} u + \bar{d} d ) | N \rangle \,,
\eeq
in which the combination $m_q (  \bar{u} u + \bar{d} d ) $ is renormalization group invariant. The scalar density is often discussed in this form as it is more directly related to the experimental quantities. 
The traditional estimate \cite{Gasser:1990ce}
is $\Sigma_N \simeq 45$ MeV (which includes up- plus down-quark contributions), but currently the estimates based on lattice QCD or combined analyses of the lattice QCD and the chiral perturbation theory range over $40-70$ MeV.
(see a review \cite{Gubler:2018ctz} and the references therein.)
Substituting $m_q\simeq 5$ MeV, we get $N_\sigma \simeq 8-14$,
so the scalar density is
\beq
 \frac{\, N_\sigma \,}{\, 4\pi R_{N}^3/3 \,}
 =\bigg( (0.24-0.30)\, {\rm GeV}\, \frac{\, 1\,{\rm fm}\,}{\, R_N \,} \bigg)^3\,,
\eeq
where $R_N$ is the size of a nucleon, which is $\sim 1$ fm. This is roughly the same order of magnitude as the vacuum scalar density but with the opposite sign.
The scalar isoscalar radius is $\langle r_s^2 \rangle \simeq ( 0.7-1.2\, {\rm fm})^2$ \cite{Schweitzer:2003sb}. 
For comparison, the values in the PDM for various $m_0$ are listed in Table.\ref{tab-sigmatermvalue}.

We note that our $m_\sigma \simeq 270-410$ MeV is smaller than the mass of the scalar meson $f_0(500)$ 
(with the width $\sim 500 $ MeV),\footnote{
It is not a trivial issue whether one should interpret $\sigma$ in mean field models as the physical scalar meson.
}
while  $\Sigma_N$ in our model is larger than the traditional value $\simeq 45$ MeV.
Recalling that $\Sigma_N \propto \partial \sigma^{\rm vac}/m_q \sim m_\sigma^{-2}$, 
better agreement with empirical values may be achieved by improving our model to accommodate a larger $m_\sigma$.
This can be done by, e.g., adding higher order polynomials of $\sigma$ fields to modify the curvature of the effective potential.
Such a fine tuning is beyond the scope of this paper;
we keep using the parameter set used for analyses of EOS in Ref.\cite{Minamikawa:2020jfj}, and shall not redo EOS construction.
We expect that overall qualitative trends of chiral restoration can be studied within the present setup.

The relation $N_\sigma = \partial m_N^{\rm vac} /\partial m_q $ tells us that the scalar charge should be positive, 
as the nucleon mass is supposed to increase  with the current quark masses. 
Since the chiral condensate in vacuum is negative, the nucleon scalar charges tend to cancel the vacuum chiral condensate.

\subsection{Dilute regime} \label{sec:dilute}

\begin{figure}[ht]
\includegraphics[width=5.2cm]{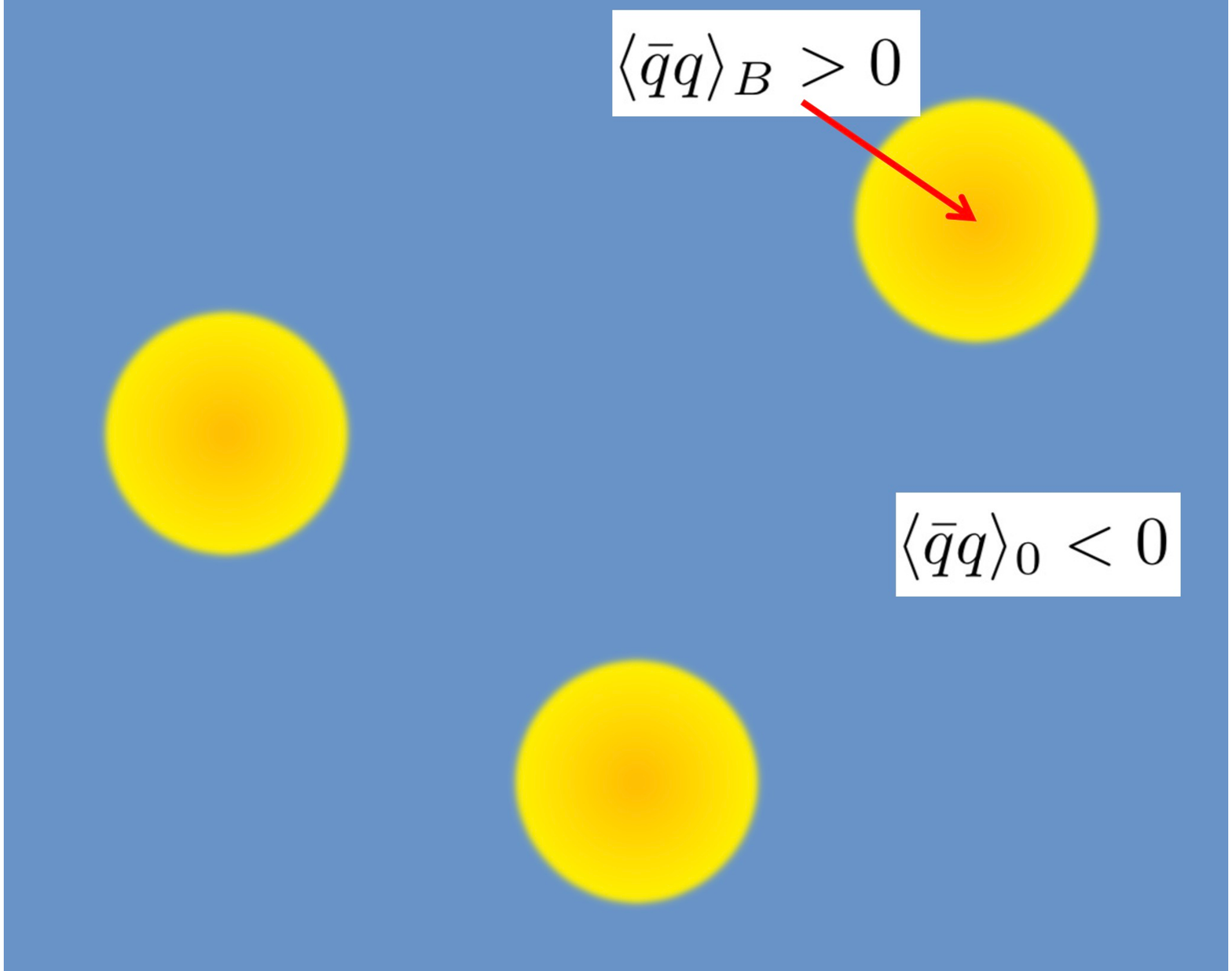}
\caption{{\small
Chiral condensates in dilute regime. The region dominated by the vacuum chiral condensate has the negative scalar charge while in nucleons the scalar charges are positive.
}}
\label{fig:chiral_in_nuclear}
\end{figure}

In the dilute regime (see Fig.\ref{fig:chiral_in_nuclear}) we can make a solid statement; here nucleons are widely separated, so the scalar density in medium is estimated by simply adding the scalar charges of nucleons to the vacuum condensate, and then taking the spatial average, i.e.,
\beq
\langle (  \bar{u} u + \bar{d} d ) \rangle \simeq \langle (  \bar{u} u + \bar{d} d ) \rangle_0 + n_B N_\sigma \,, 
\eeq
or equivalently one can write
\beq
\sigma \simeq f_\pi \bigg( 1 +  \frac{\, n_B N_\sigma \,}{\, \langle (  \bar{u} u + \bar{d} d ) \rangle_0 \,} \bigg) \,.
\eeq
This is the famous linear density approximation, which implies that the value of $\sigma$ decreases as $n_B$ increases.

\begin{figure}
\includegraphics[width=0.90\hsize]{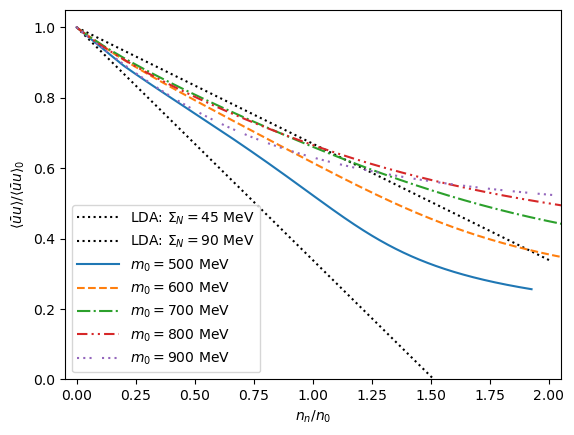}
\caption{{\small
Ratio of the quark condensate in the PDM
$\ev{\bar uu}/\ev{\bar uu}_0 = \sigma/f_\pi$
versus the neutron number density $n_n$. 
}}
\label{fig-condensate}
\end{figure}

The violation of this approximation signals the end of the dilute regime.
Figure\ref{fig-condensate} shows the ratio of the quark condensate,
$\ev{\bar uu}/\ev{\bar uu}_0 = \sigma/f_\pi$, versus the neutron number density $n_n$ in pure neutron matter 
and compare it with the linear density approximation.
Our mean field results have milder chiral restoration than in the linear density approximation. 
Similar trends have been found in the analyses based on the chiral effective theories including the fluctuations of 
pions \cite{Kaiser:2007nv,Kaiser:2008qu,Drews:2016wpi}

We stress that, in light of the PDM, the chiral restoration discussed here does not necessarily have the immediate impacts on the properties of baryons nor the Dirac sea. 
These considerations are consistent with our no sea approximation for the thermodynamic potential for nucleons (see, Eq.(\ref{eq:omega_B})) and modest changes in nucleon mass in the PDM.
We also note that, in a high temperature transition from a hadron resonance gas (HRG) to a quark gluon plasma (QGP), the chiral condensates substantially decrease just below the transition temperature, but at such temperature the HRG model with the {\it vacuum} hadron masses still work well to quantitatively describe 
the lattice data \cite{Karsch:2003zq,Andronic:2017pug} 
We will address this point again when we discuss the chiral restoration in quark matter where the modifications in the Dirac sea are likely.

\begin{table}\centering
	\caption{{\small Values of PDM in vacuum}. }\label{tab-sigmatermvalue}
	\begin{tabular}{c|ccccc}
		\hline\hline
		$m_0$ [MeV] ~&~ 500 ~&~ 600 ~&~ 700 ~&~ 800 ~&~ 900  \\
		\hline
		$(\partial m_N /\partial\sigma)_\mathrm{vac.}$ &~ 7.97 ~&~ 7.01 ~&~ 5.87 ~&~ 4.56 ~&~ 3.07 \\
		$m_\sigma$ [MeV]                                            &~ 396  ~&~ 414 ~&~ 388 ~&~ 332   ~&~ 271\\
		$\Sigma_N$ [MeV]                                            &~ 91.9 ~&~ 74.0 ~&~ 70.5 ~&~ 74.6 ~&~ 75.7\\
		\hline\hline
	\end{tabular}
\end{table}
%

\subsection{Strange quark condensate}

The increase in nuclear density also affects the strange quark condensate, as nucleons contain the sea strange quarks.
The abundance of sea strange quarks in a nucleon can be estimated by using the strangeness sigma-term,
\beq
\Sigma_{sN} = m_s \frac{\, \partial m_N \,}{\,\partial m_s} = \int_{\vec{x} } \langle N| m_s \bar{s} s | N \rangle = m_s N_s \,.
\eeq
Our PDM does not manifestly include the strangeness so that we simply substitute the vacuum value for the linear density approximation,
\beq \label{eq-strange-lda}
\langle \bar{s} s \rangle \simeq \langle \bar{s} s \rangle_0 + n_B N_s \,. 
\eeq
This estimate will be used when we consider the interpolation between nuclear and quark matter.

Our estimate of the strange quark condensate contains several errors even in dilute regime.
In fact, the strangeness content in a nucleon is difficult to determine, as it is a small effect not related to the valence quarks. 
The estimates from the lattice QCD range from $\simeq 17$ MeV to $\simeq 42$ MeV.
Taking $m_s \simeq 100$ MeV, we estimate $N_s \simeq 0.2-0.4$, which is about $\sim (0.01-0.05) N_\sigma$.

The positivity of strangeness sigma-term found on the lattice implies that $\partial m_N/\partial m_s$ is positive. 
This is by no means trivial, as strange quarks are not valence quarks in a nucleon, and increasing strange quark does not readily increase the valence quark mass nor the nucleon mass. 
One possible way to express  $\partial m_N/\partial m_s >0$ is to assume the validity of the expression as in Eq.(\ref{eq:Nsigma}),
\beq
N_s
 = \frac{\, \partial m^{\rm vac}_N \,}{\, \partial m_s \,} 
 \simeq \frac{\, \partial \sigma^{\rm vac} \,}{\, \partial m_s \,} \bigg( \frac{\, \partial m_N \,}{\, \partial \sigma \,} \bigg)_{\sigma = \sigma_{\rm vac}} \,,
\eeq
[reminder: $\sigma$ is made of the sum of up- and down-quarks, not including strange quarks]
and to consider the Kobayashi-Maskawa-'tHooft term for the $U(1)_A$ anomaly,
\beq
{\cal L}_{\rm KMT}  \sim C (\bar{u}u) (\bar{d}d) (\bar{s}s) + \cdots \,,
\eeq
where we wrote only the product of scalar densities.
A larger strange quark more explicitly breaks the chiral symmetry and hence enhances the size of strange quark condensate in vacuum.
Within mean field treatments (as in the NJL model), a larger strange condensate contributes to the term like
\beq
{\cal L}_{\rm KMT} 
 \sim 
 C  \langle \bar{s}s \rangle_0 \big[\, 
  \langle \bar{d}d \rangle_0 \times \bar{u}u 
 + \langle \bar{u}u \rangle_0 \times  \bar{d}d 
\, \big]  +\cdots \,,
\eeq
which, for $C<0$, assists the chiral symmetry breaking and increases the effective quark masses for up- and down-quarks.
These relations suggest that a larger $m_s$ firstly enhances $\langle \bar{s}s\rangle_0$, and then the effective masses of up- and down-quarks through the anomaly term. 
Within this description, $m_N$ or $\sigma_{\rm vac}$ increases for a larger $m_s$, and hence $N_s >0$ follows.

\section{Quark matter (CFL phase) } \label{sec:quark_matter}

\subsection{Overall picture} \label{sec:overall}

We use the NJL model for our quark matter descriptions.
The model does not describe the confinement, but explains the hadron phenomenology not very sensitive to the confining effects \cite{Hatsuda:1994pi}, 
e.g., the low energy constants for a hadronic effective Lagrangian for energies $\lesssim 1$ GeV. 
The NJL model is supposed to capture the physics at semi-hard scale, $0.2\, {\rm GeV} \lesssim p \lesssim 1\, {\rm GeV}$, the scale between confinement and chiral symmetry breaking \cite{Shuryak:1981fz,Manohar:1983md}. 
In terms of the distance scale, those quark models capture the physics in a hadron, $\sim 1$ fm, but do not resolve the partonic structure of constituent quarks \cite{Suenaga:2019jjv}.
We expect the NJL type constituent quark models to give reasonable descriptions on the bulk quantities at density where baryons begin to overlap.

Our NJL model includes the vector repulsion and diquark attraction in addition to the standard NJL model \cite{Hatsuda:1994pi}.
These additional interactions are motivated by the fact that the successful hadron description needs not only the use of proper constituent quark masses, 
but also the color interactions at semi-hard scale. 
For instance the color magnetic interactions are needed to account for the level splitting such as $N-\Delta$ or $\pi-\rho$ \cite{DeRujula:1975qlm}. 
We try to include the attractive part of such interactions in our diquark terms. 
At high density they lead to the diquark condensation, and, in three-flavor model, the color-flavor-locked (CFL) pairing. 
Meanwhile the magnetic interactions in repulsive channels can be used to explain the channel dependence in baryon-baryon interactions at short distance \cite{Park:2018ukx,Park:2019bsz}, 
e.g., hard core repulsion between nucleons. Our vector repulsion can be thought of the parameterization of such short distance repulsion. 

While the vector and diquark interactions contribute to the energy density with the opposite signs, their effects do not cancel much as the density dependence are different.
The vector repulsion acts on the whole bulk part of the quark Fermi sea, contributing $\varepsilon_{\rm vector} \sim + n_B^2$, while the diquark attraction mainly affects on quarks near the Fermi surface, contributing $\varepsilon_{\rm diquark} \sim - \Delta^2 n_B^{2/3}$.
Both contributions can stiffen equations of state \cite{Kojo:2014rca}.

\subsection{NJL model} \label{sec:NJL}

Our NJL Lagrangian is \cite{Kojo:2014rca}
\begin{align}\label{L_CSC}
	\lag
	&=\lag_{\rm NJL}
	+\lag_\mathrm{V}
	+\lag_\mathrm{d}
 \ .
\end{align}
The first term is the standard three-flavor NJL model with the anomaly term
($\lambda_A$ and $\tau_A$ ($A=0,\cdots 8)$ being the identity element $(A=0)$ and Gell-Mann matrices ($A=1,\cdots,8$) in color and flavor spaces, respectively, normalized as $\tr[\lambda_A \lambda_B]=\tr[\tau_A \tau_B]=2\delta_{AB}$),
\beq
	\lag_{\rm NJL} &=&\bar q( i \slash{\!\!\!\partial} -\hat m_q + \hat{\mu} \gamma_0 )q  \nonumber \\
	& +&G\sum_{A=0}^8\qty[(\bar q\tau_Aq)^2+(\bar qi\gamma_5\tau_Aq)^2] \nonumber \\
	&-&K\big[\det_f\bar q(1-\gamma_5)q+\det_f\bar q(1+\gamma_5)q \big] \ ,
\eeq
where $\hat{m}_q = {\rm diag.} (m_u, m_d, m_s)$ is the current quark mass matrix for $u,d,s$-quarks,
and $\hat{\mu}$ is
\begin{align}
	\hat{\mu}= \mu_B/3 + \mu_{c3} \lambda_3 + \mu_{c8}\lambda_8+\mu_Q Q \,,
\end{align}
with $ \mu_{c3}, \mu_{c8}, \mu_Q$ being the third and eighth components of color charges, and the electric charges;
the charge matrix is $Q= {\rm diag.} (2/3, -1/3, -1/3)$. 
For the values of the coupling constants $(G, K)$ and the UV cutoff of the model $\Lambda_{\rm NJL}$, 
we chose the values of Hatsuda-Kunihiro parameters: $G\Lambda_{\rm NJL}^2=1.835$, $K\Lambda_{\rm NJL}^5=9.29$ with $\Lambda_{\rm NJL} = 631.4$ MeV.
This set of parameters successfully describes the low energy hadron physics in vacuum, such as masses of Nambu-Goldstone bosons, decay constants, and so on. 

The vector and diquark interactions are
\beq
	\lag_\mathrm{V} &=& -g_V(\bar q\gamma^\mu q)(\bar q\gamma_\mu q) \ ,\\
	\lag_\mathrm{d} 
	&=& H\sum_{A,B=2,5,7}\left[(\bar q\tau_A\lambda_BC\bar q^t)(q^tC\tau_A\lambda_Bq)\right. \nonumber \\
	&&~~ + \left.(\bar qi\gamma_5\tau_A\lambda_BC\bar q^t)(q^tCi\gamma_5\tau_A\lambda_Bq)\right]\ , 
\eeq
where $C = i \gamma_0 \gamma_2$ is the charge conjugation matrix.
In mean field approaches, the coupling constants $(g_V, H) $ are not well constrained in vacuum as the density and diquark mean fields (see below) are vanishing.
But the impacts of these couplings are very large at finite baryon density, and the range of the couplings has been constrained by neutron star observables \cite{Baym:2019iky,Minamikawa:2020jfj,Ayriyan:2021prr} and by a model of non-perturbative gluon exchanges \cite{Song:2019qoh}.

The chiral and diquark condensates
($\tilde{\sigma}$ slightly differs from the $\sigma$ in the PDM by a factor)
\beq
\tilde{\sigma}_f  = \langle {\bar{q}_f q_f} \rangle \,,~~~ 
d_a  =  \langle q^t C\gamma_5 R_a q \rangle \,, 
\eeq
where the indices $f=1,2,3$ correspond to $u,d,s$-flavors, 
$(R_1,R_2,R_3)=(\tau_7\lambda_7,\tau_5\lambda_5,\tau_2\lambda_2)$.
The quark density $n_q (= N_c n_B) $ is the sum of $u,d,s$-quark densities
\beq
n_q = \sum_{f} n_f \,,~~~~ n_f  =   \langle q_f^\dag q_f \rangle \,.
\eeq
Within the mean field approximation, the single particle energies are obtained by diagonalizing the matrix
\beq
	S^{-1}(k) =\mqty(
	~ \slash{\!\!\! k} - \hat M+\gamma^0\hat\mu_{\rm MF} & \gamma_5 \Delta_i R_i \\
	-\gamma_5 \Delta_i^\ast R_i &  \slash{\!\!\! k} - \hat M-\gamma^0\hat\mu_{\rm MF} ~) \ , 
\label{inverse propagator}
\eeq
whose eigenvalues, $\epsilon_{i=1,..,72}$ (among which $18$ eigenvalues are independent), give the single particle (sp) energies of quarks.
The effective chiral mass, diquark gap, and effective chemical potential are 
\beq
	M_i&=&m_i-4G\sigma_i+K|\epsilon_{ijk}|\sigma_j\sigma_k\ , \\
	\Delta_i&=&-2Hd_i\ , \\
	\hat\mu_{\rm MF}&=&\mu_B/3 - 2g_Vn_q+\mu_3\lambda_3+\mu_8\lambda_8+\mu_QQ \ .
\eeq
Here $-2g_Vn_q$ is from the mean field repulsion.
The $\mu_3$ and $\mu_8$ will be tuned to enforce the color-neutrality constraints, $\partial \Omega/\partial \mu_{3,8}=0$.

With the single particle energies and condensation terms, the thermodynamic functional is given by
$\Omega[\phi] = \Omega_{\rm sp} [\phi] + \Omega_{\rm cond} [\phi] + \Omega_{\rm lepton}  $, where the quark part is
\beq
	\Omega_{\rm sp} [\phi] &=&-2\sum_{i=1}^{18}\int^\Lambda\frac{d^3\mathbf{p}}{(2\pi)^3}\frac{\epsilon_i}{2} \,, \label{Omega s}\\
	\Omega_{\rm cond} [\phi] &=&\sum_{i}(2G\sigma_i^2+Hd_i^2)-4K\sigma_u\sigma_d\sigma_s-g_Vn_q^2\,,
\eeq
and the lepton part is treated in the same way as in the PDM (Eq.(\ref{eq:lepton_EOS})); the value of $\mu_Q$ is tuned to satisfy the charge neutrality condition.
The $\phi$ in quark part collectively describes $(\sigma, d)$. After minimizing $\Omega[\phi]$ by $\phi=\phi_*$ under the neutrality constraints, 
we obtain the thermodynamic potential $\Omega =  \Omega [\phi_*]$.

\subsection{Chiral condensates in the CFL quark matter} \label{sec:chiral_NJL}

\begin{figure}[ht]
\includegraphics[width=0.8\hsize]{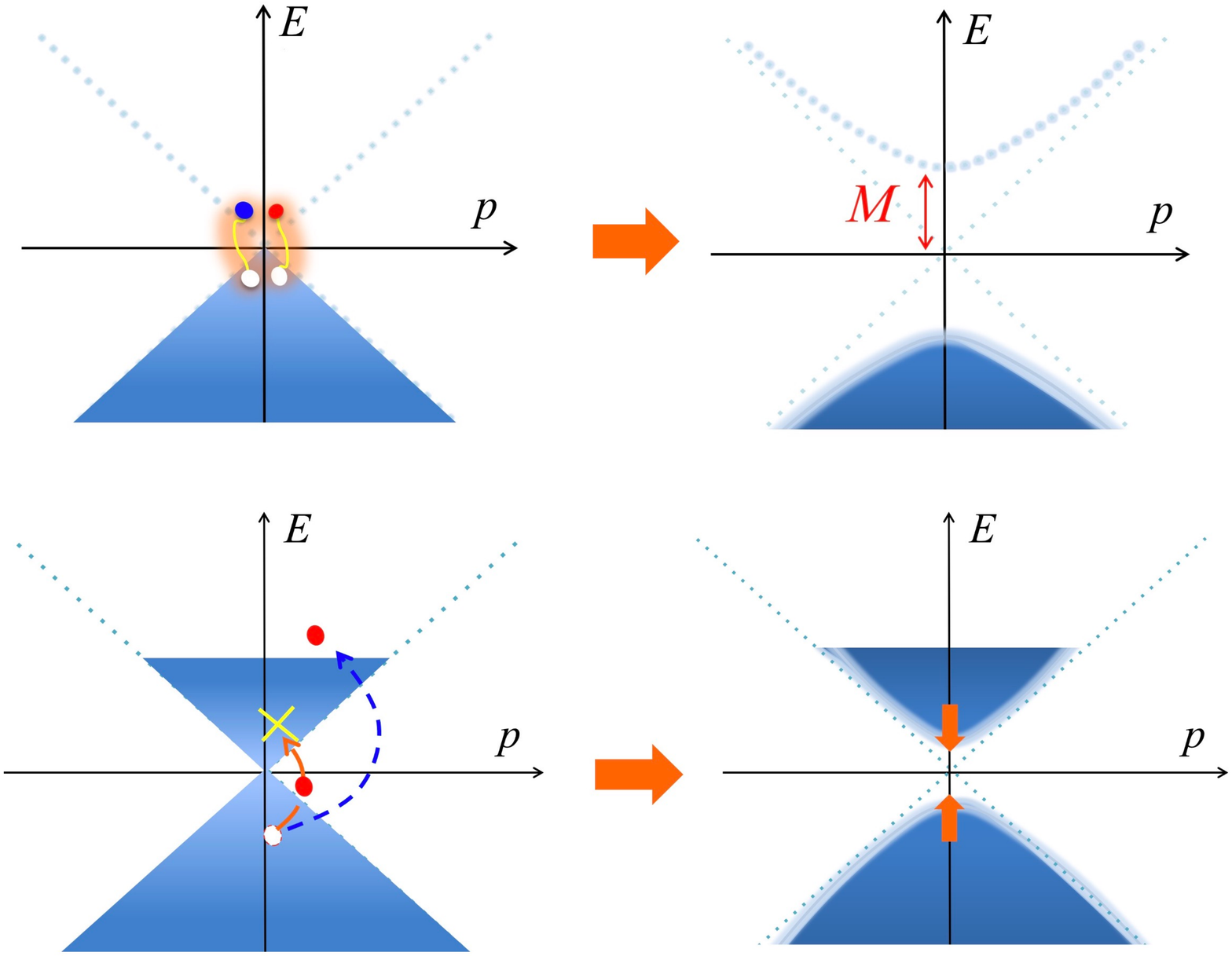}
\caption{{\small
The chiral symmetry breaking by condensation of quark-antiquark pairs; (upper) in vacuum; (lower) in medium. In the latter the pairing is blocked by the quark Fermi sea.
}}
\label{fig:chiral_in_quark}
\end{figure}

In quark matter, the quark Fermi sea disfavors the quark-antiquark pairing which triggers the formation of the chiral condensates (Fig.\ref{fig:chiral_in_quark}). 
This is because having an antiquark costs the energy of the order of the quark Fermi energy; 
regarding an antiquark as a hole in the Dirac sea, its creation needs to bring the particle in the Dirac sea to the domain beyond the Fermi sea, avoiding the Pauli blocking. 
Hence, as density increases, the chiral condensates naturally dissociate. 
Alternatively, the particle-particle or particle-hole pairings are favored as they involve only the degrees of freedom near the Fermi surface \cite{Kojo:2009ha}.

It is important to note that the chiral restoration discussed here differs from what we have described in dilute nuclear matter;
there the chiral restoration is due to the cancellation between positive and negative charges. 
But the positive scalar charge in each nucleon nor the negative scalar charges from the vacuum do not change much with density; 
only the sum does. Meanwhile, in quark matter we do consider the reduction in each of positive and negative scalar charges; 
both the Dirac sea and the constituent quark masses change. 
In the context of the PDM, the Dirac sea modification should be responsible for the reduction of the chiral invariant mass $m_0$.

\section{Condensates in a unified EOS } \label{sec:unified_EOS}

In this section we construct a unified EOS and condensates by interpolating the nuclear and quark models in the previous sections.
We compute the chiral and diquark condensates, and the composition of matter with ($u,d,s$)-quarks and leptons (electrons and muons, $e,\mu$).


\subsection{Unified generating functional}

In order to compute a unified EOS and a condensate $\phi$, we first construct a generating functional $P(\mu_B; J)$ in which $J$ is the external field coupled to the condensate $\phi$. 
After computing the generating functional, one can differentiate it with respect to $J$,
\beq
\phi = - \frac{\, \partial P \,}{\, \partial J \,} \bigg|_{J=0} \,,
\eeq
to calculate the condensate at a given $\mu_B$.

The generating functional for the nuclear domain, $n_B \leq 2n_0$, is given by the PDM, and for the quark matter domain, $n_B \geq 5n_0$, by the NJL model.
These functionals are interpolated by requiring the continuity up to the second derivatives at the boundaries, $2n_0$ and $5n_0$.
As an interpolating function we use a polynomial function with six constants $a_n(J)$,
\begin{align}
P_\Interp(\mu_B ; J)&=\sum_{n=0}^5a_n(J)\mu_B^n\,,
\end{align}
for the intermediate region $2n_0<n_B<5n_0$. 
We write the chemical potentials at the boundaries as $\mu_B^L$ and $\mu_B^U$ which satsify 
\beq
n_B (\mu_B^L; J) = 2n_0 \,,~~~~~ n_B (\mu_B^U; J) = 5n_0 \,.
\eeq
It is important to remember that, as we use the fixed densities for the boundaries, $\mu_B^L$ and $\mu_B^U$ depend on $J$.
The six parameters $a_n$ are determined from the six boundary conditions,
\beq
\frac{\, \partial^k P_\Interp \,}{\, ( \partial \mu_B )^k \,} \bigg|_{\mu_B^L (\mu_B^U) }
	= \frac{\, \partial^k P_{\rm PDM (NJL)} \,}{\, ( \partial \mu_B )^k \,} \bigg|_{\mu_B^L (\mu_B^U) } \,,
\label{eq:matching}
\eeq
where $k=0,1,2$. Determination of $a_n$ at $J=0$ gives us the unified EOS.

The model parameters of the generating functional are restricted by the causality condition, 
\begin{align}
c_s^2&=\dv{P}{\varepsilon}=\frac{n_B}{\mu_B\chi_B}\leq1\,, 
\end{align}
where $c_s$ is the sound velocity 
and $\chi_B=\partial^2P/\partial\mu_B^2$ is the baryon number susceptibility.

\subsection{A practical method to compute condensates} \label{sec:practical}

In this subsection, we explain a practical method to compute condensates from our generating functional.
In this method, it is not necessary to manifestly compute $P(\mu_B,J)$ for various $J$, but we utilize only the $\mu_B$-dependence of the condensate at $J=0$ for each interpolating boundary.
This method is useful especially when we need to compute many condensates.

For our interpolating function, the expression of condensate $\phi$ in the interpolated domain is given by
\begin{align}
\phi_\Interp =-\pdv{P_\Interp }{J}\bigg|_{J=0}=-\sum_{n=0}^5{\pdv{a_n}{J}}\bigg|_{J=0}\mu_B^n\,. 
\label{eq:inter_condensate}
\end{align}
Thus the determination of $\phi_\Interp$ is equivalent to the determination of six constants $\partial a_n/\partial J$ ($n=0,...,5$) at $J=0$.
Recalling that our matching condition at a given $J$ is given in Eq.(\ref{eq:matching}),
we obtain the $\partial a_n/\partial J$ by taking the $\mu_B$-derivatives of Eq.(\ref{eq:matching}),
%
\beq \label{eq-dphidmu}
\frac{\partial}{\, \partial J \,} \bigg( \frac{\, \partial^k P_\Interp \,}{\, ( \partial \mu_B )^k \,} \bigg|_{\mu_B^L (\mu_B^U) } \bigg)
	= \frac{\partial}{\, \partial J \,} \bigg( \frac{\, \partial^k P_{\rm PDM (NJL)} \,}{\, ( \partial \mu_B )^k \,} \bigg|_{\mu_B^L (\mu_B^U) } \bigg) \,,
	\nonumber\\
\eeq
%
%
where $k=0,1,2$. We set $J=0$ in the end.
When we compute the derivatives,
we should remember that $\mu_B^L (\mu_B^U)$ depend on $J$, see Appendix.~\ref{sec-detailedcalculations} for the details.
Importantly, all these derivatives at $J=0$ can be obtained using only quantities at a given $\mu_B$ and $J=0$; it is not necessary to refer to the quantities for various $J$ nor $\mu_B$; 
our computations are {\it local}.
Hence our method reduces computational efforts significantly.

\subsection{Numerical results}

\begin{figure}[ht]
\includegraphics[width=0.9\hsize]{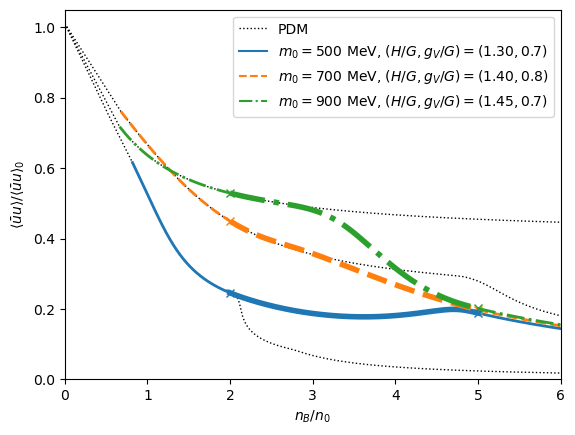}
\caption{{\small
Density dependence of the ratio of the chiral condensates. 
}}
\label{fig-interpolate-qqbar}
\end{figure}

Using the method presented in the last section,
we compute the light quark chiral condensate $\ev{ (\bar{u}u + \bar{d} d) }$, 
the strange quark condensate $\ev{\bar{s}s}$, 
the diquark gaps $\Delta_i$, and the quark number densities $n_f$,
from nuclear to quark matter domain. 
Unless otherwise stated we pick up three sample parameter sets for $(m_0\, [{\rm MeV}], H/G, g_V/G) $ as
\beq
&& (500, 1.30, 0.7) \,, ~~~~[{\rm blue~solid}]  			\nonumber \\
&& (700, 1.40, 0.8) \,, ~~~~[{\rm orange~dashed}]  		\nonumber \\
&& (900, 1.45, 0.7) \,, ~~~~[{\rm green~dash~dotted}]
\eeq
all of which lead to EOS with the causal speed of sound.
Here [...] indicates the types of lines used in figures for these parameters. 
As a guide, we will also show the extrapolation of the PDM results by black dotted lines.

\subsubsection{Light quark chiral condensates} \label{sec:light_quark_condensate}

Figure\ref{fig-interpolate-qqbar} shows 
the ratio of the chiral condensate in medium to the vacuum counterpart, 
$\ev{ (\bar{u}u + \bar{d} d)} /\ev{ (\bar{u}u + \bar{d} d) }_0$.
It is clear that the condensate at the boundaries put strong bias to the condensate in the interpolated domain.

The PDM results are sensitive to the choice of $m_0$.
For $m_0 =500$ MeV, the chiral condensate reduces radically, 
as the $N$-$\sigma$ coupling must be strong to reproduce $m_N = 939$ MeV, 
and hence changes in nuclear medium have large impacts on the behavior of $\sigma$.
For a larger $m_0$, the medium effects on $\sigma$ are smaller.
Meanwhile the NJL model typically leads to the chiral condensates whose magnitudes are $20-30\%$ of the vacuum values.

As we have mentioned in Sec.\ref{sec:dilute}, the chiral restoration within the PDM may underestimate or neglect the chiral restoration at the quark level.
At higher density the effects of interactions among nucleons are stronger, 
and it is natural to consider structural changes of nucleons and the modifications of $m_0$ and the couplings.
The bias from the quark matter side is used to infer the trends of these effects.
Therefore, although phenomenological, we believe that our interpolating method offers reasonably balanced descriptions.

\subsubsection{Strange chiral condensates} \label{sec:strange_condensate}

\begin{figure}[ht]
\includegraphics[width=0.9\hsize]{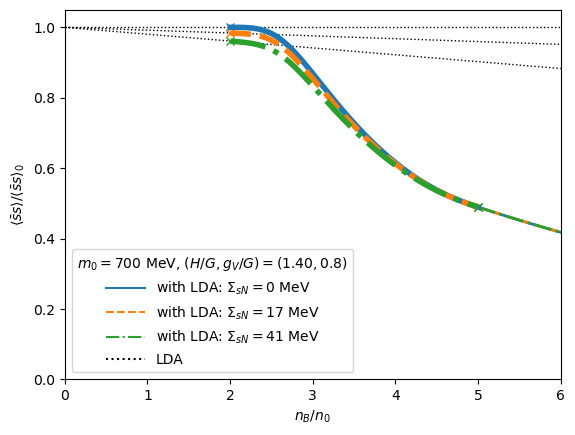}
\caption{{\small
Density dependence of the strange quark condensate normalized by the vacuum counterpart. 
}}
\label{fig-interpolate-ssbar}
\end{figure}

Another quantity of interest is the strange quark condensate, as shown in Fig.\ref{fig-interpolate-ssbar}.
Let us first look at the trend at $n_B \leq 2n_0$.
Although the strange quark contributions are not defined in the PDM of this work, 
we estimate them by assuming the linear density approximation Eq.(\ref{eq-strange-lda}) with the strangeness sigma-term. 
For the sigma-term we use $\Sigma_{sN} =0,17, 41$ MeV. 
For $\Sigma_{sN}=0$, the strange quark condensate is density independent, 
while the values $\Sigma_{sN}=17,41$ MeV are picked up from Ref.\cite{Gubler:2018ctz}. 
We fixed the parameters $m_0=700$ MeV and $(H, g_V)/G=(1.4,0.8)$, and then vary the values of $\Sigma_{sN}$.
As expected, Fig.\ref{fig-interpolate-ssbar} shows that the chiral restoration in the strange quark sector
is very small in the nuclear domain.

Beyond $2n_0$, the strange quark condensate starts to reduce, and the reduction from the vacuum value is $\sim 20\%$ at $n_B \simeq 3n_0$.
By construction, this reduction is due to the bias from the boundary condition in the quark matter side.
In the quark matter at $n_B \simeq 5n_0$, the reduction is $\simeq 50\%$.
This chiral restoration should mainly come from two effects.
One is the suppression of the anomaly term, $\sim \langle \bar{u}u \rangle \langle \bar{d}d \rangle (\bar{s}s)$, associated with the reduction of light quark condensates.
The other is due to the appearance of the strange quark Fermi sea.
In our model the strangeness density begins to share $\sim 10\%$ of quark density at $n_B \simeq 3n_0$, 
where we suppose that the strange quark sector is modified at the quark level.

\subsubsection{Diquark gaps and number density} \label{sec:diquark_condensate}

\begin{figure}[ht]
	\includegraphics[width=0.9\hsize]{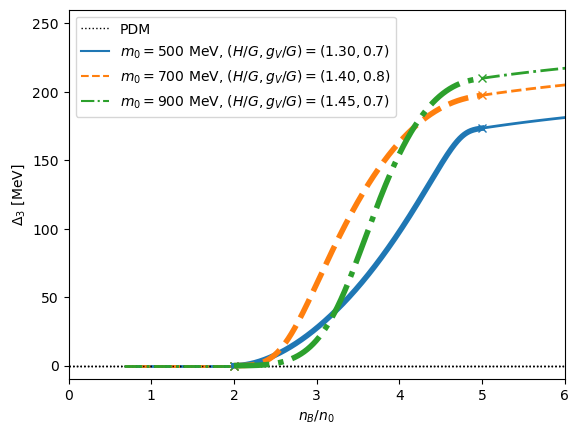}
	\includegraphics[width=0.9\hsize]{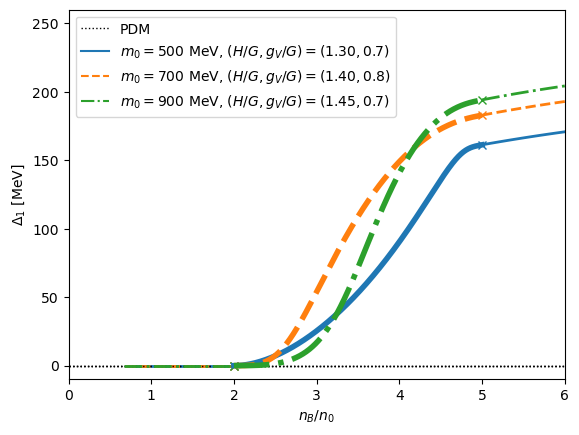}
\caption{{\small 
Diquark condensates as functions of density; (upper) $\Delta_3 = \Delta_{ud}$, and (lower) $ \Delta_1 = \Delta_{ds}$.
}}
\label{fig:diquarks}
\end{figure}

\begin{figure}[ht]
	\includegraphics[width=0.9\hsize]{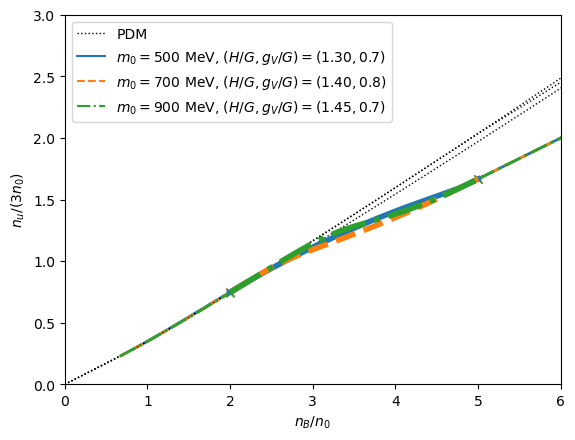}
	\includegraphics[width=0.9\hsize]{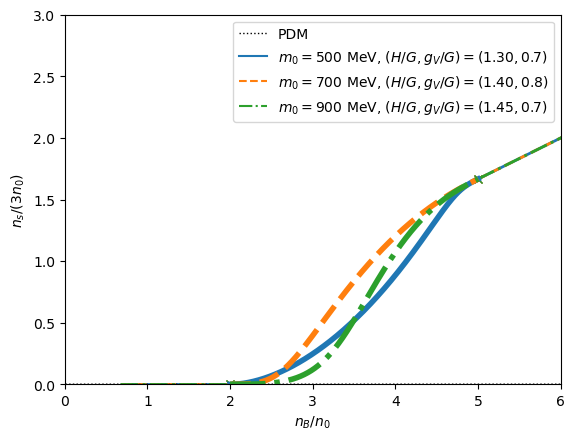}
\caption{{\small 
Number density for (upper) up-quarks, and (lower) strange-quarks.
}}
\label{fig:number}
\end{figure}

Figure\ref{fig:diquarks} shows the numerical result of the interpolated diquark gaps 
in the $ud$-pairing channel (upper panel)  and $ds$-pairing channel (lower panel)
for the same choice of parameters in Fig.\ref{fig-interpolate-qqbar}. 
The diquark condensates are assumed to be zero at $n_B \le 2n_0$. 
In the CFL quark matter, the isospin symmetry holds in very good accuracy,
while in nuclear matter the pairing is zero, so
$\Delta_{ds} \simeq \Delta_{us}$ hold in good accuracy for whole domain.

Next we examine the correlation between the diquark condensates and quark number density shown in Fig.\ref{fig:number}. 
In the nuclear domain, the quark density for each flavor is calculated from the proton and neutron densities ($n_p$ and $n_n$) as
$n_u=2n_p+n_n$, $n_d=n_p+2n_n$, and $n_s=0$. 
We note that the growth of the diquark condensates is strongly correlated with the growth of quark number density,
as can be seen from the comparison of Figs. \ref{fig:diquarks} and \ref{fig:number}.
This is qualitatively reasonable;
a larger Fermi surface allows more diquark pairs,
and the associated energy reduction of the system in turn assists the growth in density.
Another important effect is on the flavor asymmetry.
In particular, the number density of the strange quark in the nuclear matter is zero, while 
the strong pairing between the strange quark and the other light quarks favor the equal population of $u,d,s$-quarks in the quark matter. As a result the number density of the strange quark rapidly increases in the interpolated region and 
the onset of the strangeness takes place at lower density than without the pairings.
This competition between the mass asymmetry and the pairing effects determines the fraction of each quark.

\subsubsection{Compositions} \label{sec:fraction}

\begin{figure}[ht]
	\includegraphics[width=0.85\hsize]{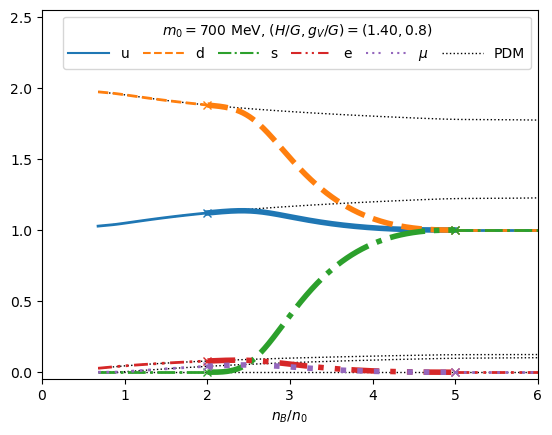}
\caption{{\small 
Matter composition $n_f/n_B$ ($f=u,d,s$) and $n_l/n_B$ as functions of baryon density.
}}
\label{fig:udsl}
\end{figure}

Finally we examine the composition of matter including leptons.
Figure\ref{fig:udsl} shows the fraction of quark density for each flavor and lepton fraction per baryon density, 
\beq
\frac{\, n_{f} \,}{\, n_B \,}\,,~~~ \frac{\, n_{l} \,}{\, n_B \,}\,, 
\eeq
where $f=u,d,s$, and the subscript $l$ refers to electron or muon densities.
The lepton fraction is determined by the charge neutrality and $\beta$-equilibrium condition, which is controlled by the charge chemical potential $\mu_Q$.

One of distinct features in our unified model is the evolution of the strangeness fraction.
The strangeness fraction becomes significant around $n_B\simeq 2.5n_0$, and strange quarks are as abundant as up- and down-quarks at $n_B \gtrsim 4.5n_0$.
At $n_B = 5n_0$, the matter becomes the CFL quark matter where the sum of quark densities satisfies the charge neutrality condition; no leptons needed in the CFL phase.
This trend with the strangeness differs from pure nuclear models in which lepton fraction increases as the baryon density does.

\section{ Discussions } \label{sec:discussions}

In this work, the chiral condensates in the crossover domain are based on the interpolation of those in nuclear and quark matter domains, 
so our pictures for the chiral restoration are somewhat indirect.
Especially the role of confinement has been obscure.
In this section we conjecture several scenarios to infer the microphysics in the interpolated domain.

\subsection{ Casher's argument and the chiral scalar density in nucleons} \label{sec:Casher}

For the relation between confinement and chiral symmetry breaking, Casher  suggested 
that the absence of the chiral symmetry breaking effects do not allow the descriptions of quark confinement  \cite{Casher:1979vw}. 
In our arguments we slightly relax Casher's arguments by just assuming the presence of chiral variant fields $(\sigma, \vec{\pi})$. 
The overall size of the chiral variant fields is characterized by
$\phi^2 \equiv (\sigma^2 + \vec{\pi}^2) $ which is chiral invariant, 
while the direction of the four vector $(\sigma, \vec{\pi})$ is chiral variant.

\begin{figure}[ht]
\vspace{-1.5cm}
\includegraphics[width=8.5cm]{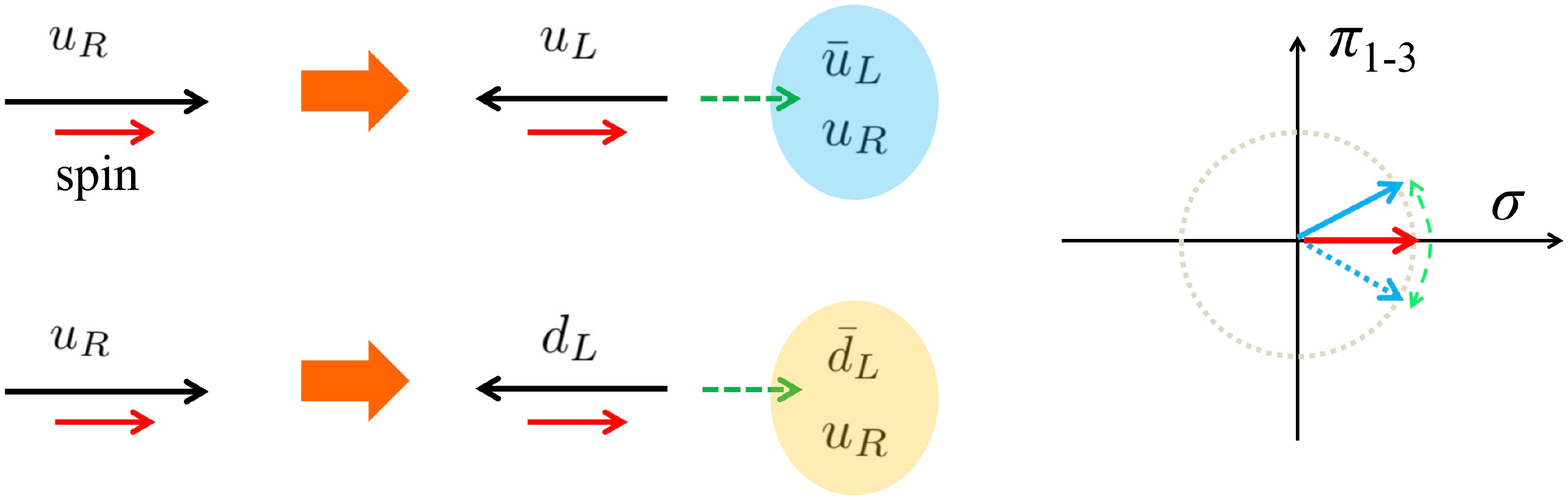}
\vspace{-1.2cm}
\caption{{\small
Helicity and spin conservation due to the emergence of the ($\sigma,\vec{\pi}$) fields.
The confinement turns a right-moving quark, $u_R$, into a left-moving quark, $u_L$ ($d_L$), but without changing the spin.
The helicity is conserved by fields made of $\bar{u}_L u_R$ ($\bar{d}_L u_R$).
The ($\sigma,\vec{\pi}$) fields fluctuate whenever the chirality of quarks flip.
}}
\label{fig:Casher}
\end{figure}

The Casher's argument focus on the helicity for a massless quark. 
The confinement requires the massless quarks to change the directions, but it does not flip the spin, violating either the conservation of the helicity or angular momentum.
To avoid such violation the confining boundaries must develop 
fields\footnote{
One can also think of the instantons as the sources of the chirality flipping; in this picture $\sigma$ appears due to quarks bound to instantons \cite{Schafer:1996wv}.
 }
 which can carry the quanta of quarks just before the reflection, see Fig.\ref{fig:Casher}.
The candidates of such fields are $\sigma$ and $\vec{\pi}$ which transform a left-(right-)handed quark into a right-(left-)handed one,
and at the same time the four vector $(\sigma, \vec{\pi})$ rotates to conserve the helicity.
(This intuitive picture will be further developed in the next section.)

Now let us apply this picture and see what consequences would follow.
Firstly the Casher's argument naturally leads to the chiral scalar density in hadrons.
This may be related to the so-called in-hadron condensate \cite{Brodsky:2010xf}. 
Meanwhile, to phrase the vacuum condensates in this context, one should first consider the $\sigma$ meson as a confined particle, 
and then argue the condensation of the $\sigma$ mesons.
Once the space is filled with many $\sigma$, the quarks can flip the chirality anywhere.

\subsection{ Topology of pion clouds and spatial modulations of the chiral scalar density} \label{sec:topology}

\begin{figure}[ht]
\vspace{-1.0cm}
\includegraphics[width=8.0cm]{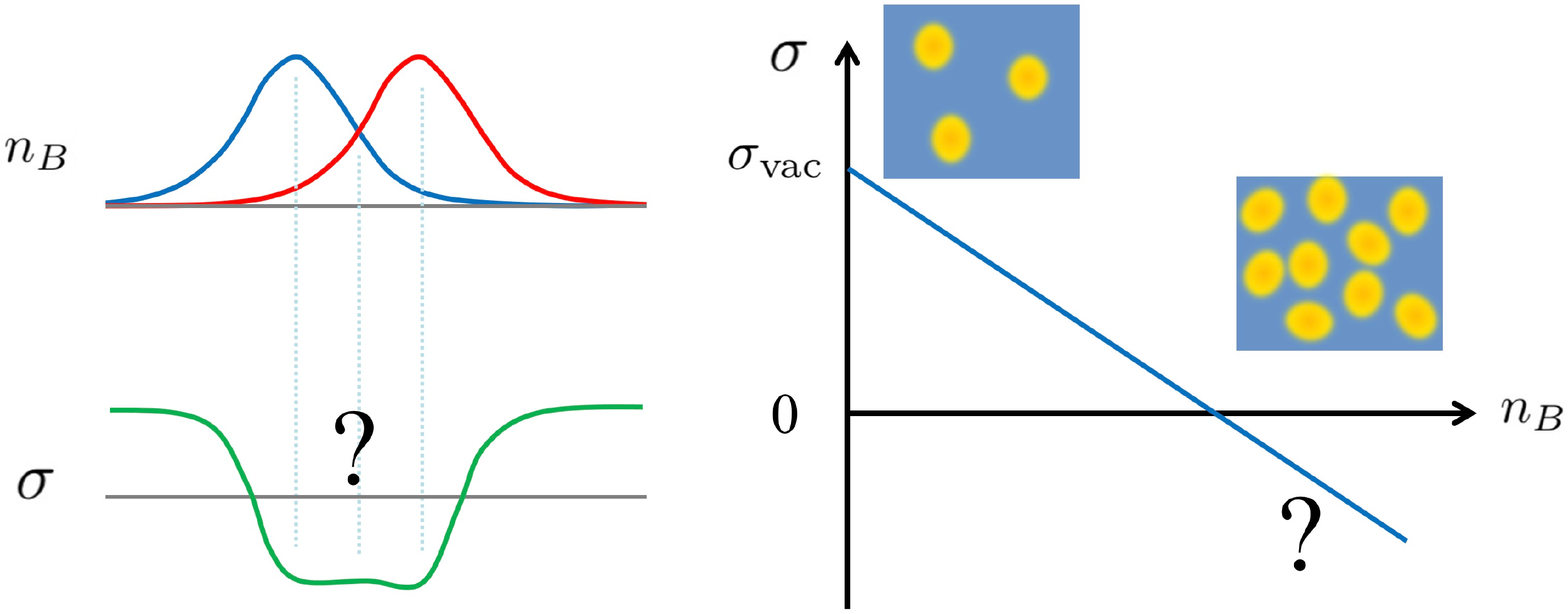}
\vspace{-1.2cm}
\caption{{\small
Naive extrapolation of the linear density approximation for the chiral condensate. As baryons overlap, the positive chiral scalar density (negative $\sigma$) from baryons
dominates over the negative scalar density from vacuum one ($\sigma >0$).
}}
\label{fig:naive_chiral}
\end{figure}

The nucleon scalar density has the sign opposite to the vacuum chiral condensate but with the similar magnitude.
Extrapolating this picture would lead to the expectation that, as nucleons overlap, they form a large domain with the positive chiral scalar density (negative $\sigma$), 
dominating over the negative chiral scalar density (positive $\sigma$) from the vacuum, see Fig.\ref{fig:naive_chiral}.
This consideration does not smoothly match with quark model descriptions at high density, where the chiral scalar density approaches zero, rather than the positive value.
To understand this discrepancy, below we first consider a two baryon system and chiral scalar density in it.

\begin{figure}[ht]
\vspace{-1.5cm}
\includegraphics[width=0.9\hsize]{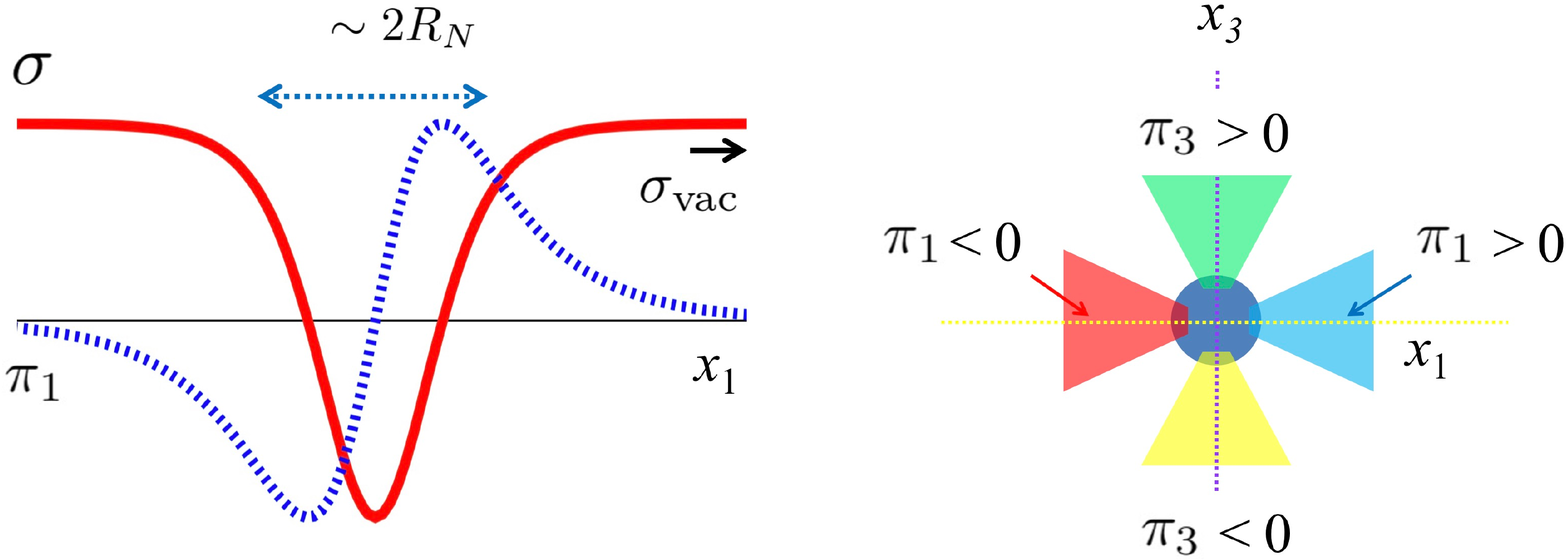}
\vspace{-1.2cm}
\caption{{\small
The expected behavior of chiral scalar ($\sigma$) and pseudo scalar fields ($\vec{\pi}$) near a nucleon (one dimensional slice).
The chiral invariant combination of these fields are $\sigma^2 + \vec{\pi}^2 \simeq$ const.
The right is the hedgehog profile.
}}
\label{fig:sigma_pi_baryon}
\end{figure}

\begin{figure}[ht]
\vspace{-0.5cm}
\includegraphics[width=8.5cm]{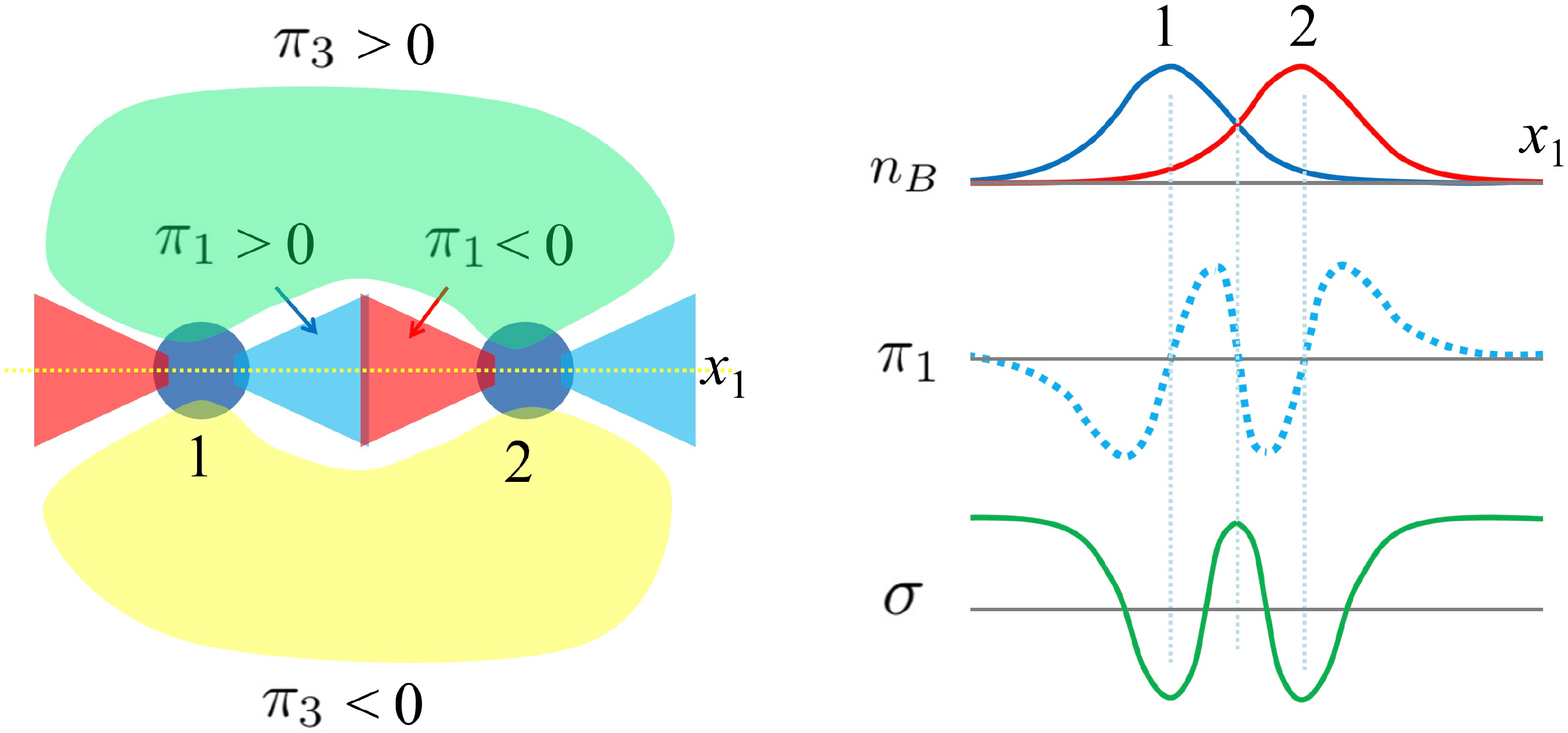}
\vspace{-1.2cm}
\caption{{\small
Two nucleons at close distance, with the hedgehog pion configurations, where the vectors $\vec{\pi}$ and $ \vec{x}$ are parallel.
The rightmost part shows the baryon density, $\pi_1$ fields, and $\sigma$ fields along the $x_1$ axis.
The negative $\sigma$ fields are accumulated at the center of baryons, while the positive $\sigma$ are accumulated at the interface of two baryons.
}}
\label{fig:two_soliton}
\end{figure}

For this purpose we first examine in more detail how the $\sigma$ field changes the sign from the vacuum to the inside of a nucleon.
Considering that the effective Lagrangian is chiral symmetric, the effective potential is a function of $\phi^2 $. 
It takes a minimum at $\phi = \sigma_{\rm vac}$ for the vacuum.
We suppose that the field variation along this circle ($\phi = {\rm const}.$) 
does not cost much energy.
Then the $\sigma$ around a nucleon should accompany $\pi$ fields whose magnitudes are large near the surface of nucleons, and are vanishing at the center, as shown in Fig.\ref{fig:sigma_pi_baryon}.
The spatial average of $\pi$ fields is zero.
If we arrange isospin distributions for these $\vec{\pi}$ fields properly to generate the topological number one as in the hedgehog form, we arrive at descriptions 
similar to the chiral soliton models \cite{Skyrme:1961vq,Adkins:1983ya,Diakonov:1987ty,Manohar:1984ys,Kahana:1984dx,Zahed:1986qz,Hata:2007mb,Nawa:2006gv}.

The topological numbers of pions around nucleons give important constraints on how baryons come close together. 
Let us consider two nucleons which are close in distance (Fig.\ref{fig:two_soliton}).
As we mentioned, in a naive description two domains of the positive scalar density simply merge to form one large domain with the positive scalar density
(negative $\sigma$) as was shown in Fig.\ref{fig:naive_chiral}. 
This picture is corrected by considering the topological constraint as shown in Fig.\ref{fig:two_soliton}. 
For two nucleon problems, the configuration of $(\sigma, \vec{\pi})$ must have nodes to generate topological numbers 
two\footnote{In (1+1)-dimensional models such statements become exact by applying the bosonization method \cite{Witten:1983ar,Witten:1978qu,Affleck:1985wa,Affleck:1985wb}; $n_B \sim \partial_1 \varphi$, $\sigma \sim \cos \varphi$, and $\pi \sim \sin \varphi$ (for $U(1)$).
Studies of dense systems can be found, e.g. in Refs.\cite{Schon:2000he,Bringoltz:2006pz,Kojo:2014fxa,Kojo:2011fh}.
}, instead of zero.
Hence $(\sigma, \vec{\pi})$ should have spatial modulations with positive and negative scalar densities, instead of forming single large domain with the positive scalar charge.
As more and more nucleons are packed, the spatial modulations of $(\sigma, \vec{\pi})$ fields become finer.
The positive and negative $\sigma$ fields tend to cancel only in the spatial average.
The magnitudes of the modulations are controlled by the size of $\phi$, not by the value of $\sigma$.
As far as $\phi$ is nonzero, it is natural to expect that the nucleon and constituent quark masses are substantial, but still we can describe the chiral restoration in the sense of $\int_{\vx} \sigma (x) = 0$.
This picture fits to the concept of the chiral invariant mass in the PDM which can be nonzero even for $\sigma = 0$
as discussed in Refs.\cite{Ma:2013ooa,Harada:2015lma}.

In the above argument we address the possibility that 
each nucleon accompanies the spatial modulations in $\sigma$ and $\vec{\pi}$ in a topological configuration.
(For chiral density waves in the PDM, see Refs.\cite{Abuki:2018ijb,Takeda:2018ldi,Heinz:2013hza}. )
In compressing nuclear matter, such modulations do not cancel, but just get squeezed altogether as far as $\phi$ remains finite.
This picture shares several viewpoints with the scenarios of the soliton crystals \cite{Klebanov:1985qi,Rho:2009ym,Kim:2007vd,Forkel:1989wc,Nawa:2008uv} or chiral spirals \cite{Buballa:2014tba,Deryagin:1992rw,Nickel:2009wj,Carignano:2010ac,Rapp:2000zd,Nakano:2004cd,Kojo:2009ha,Kojo:2010fe,Kojo:2011cn,Pisarski:2018bct,Pisarski:2020dnx};
if the compressed nucleons favor the periodic structure, they also lead to the lattice in ($\sigma, \vec{\pi}$).
Such inhomogeneous chiral condensates have been discussed in both nucleonic models and quark models.
Whether they can connect by the quark-hadron continuity is an interesting question, 
see Ref.\cite{Ma:2016gdd} and references therein (Refs. cited) for instance.

\subsection{ Diquarks } \label{sec:diquarks}

Finally we consider how diquark condensates develop from nuclear to quark matter.
As we have assumed that nucleon fields (before the diagonalization of the mass matrix) transform as $(1/2, 0)_L \oplus (0,1/2)_R$ in terms of the chiral multiplet, 
a simple way to combine three quarks is to first consider diquarks in the color antitriplet representation and then put a leftover quark. The diquarks 
\beq
[ u_L d_L ]_{I=0} \,,~~~~~[ u_R d_R ]_{I=0} \,,
\eeq
are $SU(2)_L$ and $SU(2)_R$ singlet, respectively, and hence are invariant under the chiral transformations for $SU(2)_L \otimes SU(2)_R$.
We attach the leftover quark to these diquarks, then the resulting nucleon follows the chiral transformation of the leftover quark.
Below we further assume that the diquark takes $J^P = 0^{+}$ and the spatial S-wave, for which the attractive correlations are the largest.
In this respect, the nucleon already contains the source of diquark condensates for the high density matter.

At low density the diquark correlations may exist but a diquark is tightly bound to a leftover quark to form a nucleon.
In this regime the diquarks are hidden inside of nucleons; the diquark fields may be associated with nucleons, but they do not take the form of condensates which spread over the system.
The bulk properties of the system are described in terms of nucleons.
The trend begins to change when quark exchanges between nucleons become very frequent.
Here diquarks are available not only as constituents of nucleons, but as participants to the bulk properties of the system.
As the diquark fields overlap, they develop the coherence; the phase of condensates, $\varphi$ in $\Delta  =|\Delta | {\rm e}^{ i \varphi }$, correlate over long distance.
In these pictures, the diquark condensates begin to develop when nuclear many-body forces become important ($n_B \gtrsim 2n_0$), gradually develop toward quark matter, and get established at high density 
($n_B \gtrsim 5n_0$)\footnote{
The onset of diquark pair condensations may be constrained by NS cooling.
In particular, some cooling curves disfavor substantial CFL cores which would induce too rapid cooling \cite{Blaschke:1999qx,Grigorian:2004jq}. 
One possible interpretation is that such observed NSs are too light and do not have large core densities.
But if those NSs turn out to be heavy and have large core densities, they put stringent constraints on the CFL phase.
For cooling of accreting NSs, see Ref.\cite{Cumming:2016weq}.
}.

Based on the same picture, we expect that diquark fields with strangeness are introduced to the system as the constituents of hyperons. 
Then the diquark fields gradually develop as the hyperons dissociate through the quark exchanges.

%
%


\section{ Summary} \label{sec:summary}

In this paper, we have elaborated a recipe to compute various condensates in the domain between nuclear and quark matter, assuming the quark-hadron continuity picture.
After looking at the behaviors of these condensates,
we then conjecture qualitative scenarios which interpolate nuclear and quark matter descriptions.

We found that the PDM with the substantial chiral invariant mass, $m_0 \gtrsim 500$ MeV, 
has a number of favorable properties in descriptions for EOS (as found in the previous works) and also for various condensates.
In particular the substantial reduction of $\sigma$ in dilute regime (driven by the positive scalar charge in nucleons) does not change the nucleon properties drastically. 
This is consistent with our neglect of the nucleon Dirac sea and use of fixed nucleon-meson couplings for $n_B \lesssim 2n_0$.
In our view, the intrinsic properties of nucleons begin to substantially change at $n_B \gtrsim 2n_0$,
where quark exchanges among baryons become frequent; since baryons are made of quarks, the quark exchanges are supposed to change the baryon structure.
The quarks, which are partially released, are also affected by the medium and should change the properties such as the effective mass.
The direct descriptions of such changes is difficult, but at least we can constrain it through the quark matter constraints at high density.
Our interpolation scheme is a practical way to implement these ideas.

This work has addressed only few aspects on the chiral symmetry in dense matter. There remain many issues to be addressed. Here we list up some:

(i) The PDM model can be extended to include hyperons
\cite{Chen:2009sf,Chen:2010ba,Chen:2011rh,Nishihara:2015fka,Dexheimer:2012eu,Motornenko:2019arp}. 
For NS matter, the charge chemical potential $\mu_Q$ ranges from $-100$ MeV to $-200$ MeV around $n_B \sim 1-2n_0$,
and hyperons may appear for $2-3n_0$, not far from our choice for the hadronic boundary $n_B=2n_0$ (see for instance Sec.III in Ref.\cite{Kojo:2020ztt}).
The slight extension from $n_B=2n_0$ to higher density and manifest treatment of hyperons 
would give more concrete descriptions of the strangeness than in this work.

(ii) The detailed understanding of a nucleon and its meson cloud should give a guide on the chiral symmetry in dense matter.
An important question in the context of the PDM is how a $(\sigma, \vec{\pi})$ cloud differs for the positive ($N(939)$) and negative parity ($N(1535)$) nucleons.
In terms of the constituent quark models, $N(1535)$ contains the P-wave excitation of a quark, with the larger spatial size than $N(939)$.
How this size scale estimate and $\sigma$ in the PDM are related is an important question to understand the medium effects in the PDM.
In general the medium effects should influence hadrons with the larger size, as they have closer in distance to the other hadrons. 

(iii) Ultimately our patchwork of nuclear and quark matter descriptions should be replaced with a description based on a single model.
Baryons should be constructed explicitly in terms of quarks.
Several recent works \cite{Fukushima:2020cmk,McLerran:2018hbz,Jeong:2019lhv,Kojo:2021ugu}, although schematic, have given concrete descriptions of quarks from nuclear to quark matter domains.
One of the important outcome is the peak in the speed of sound \cite{McLerran:2018hbz,Jeong:2019lhv,Kojo:2021ugu} which has been a puzzling feature inferred from NS observations \cite{Tews:2018kmu}. 
But detailed questions such as the fate of chiral symmetry breaking or diquark correlations have not been addressed in such modelings.
Our descriptions in this paper should give some hints for the fuller understanding of nuclear-quark matter phase transitions.

\begin{acknowledgments}
The work of T.M. and M.H. was supported in part by JSPS KAKENHI Grant No. 20K03927. 
T.M. was also supported in part by the Department of Physics, Nagoya University. 
T.K. was supported by NSFC Grant No. 11875144.
\end{acknowledgments}

\appendix

\section{Calculation of $\partial a_n/\partial J$ in Eq.(\ref{eq:inter_condensate})}\label{sec-detailedcalculations}

We write $\vec{P}$ as the vector which has the components of 
the six values of $\partial^k P/(\partial \mu_B)^k$ ($k=0,1,2$) 
at the boundaries $\mu_B=\mu_{B}^L,\mu_{B}^U$ 
calculated from the PDM and the NJL model. 
Since $P_I$ is a polynomial of $\mu_B$, 
the vector of the values of $\partial^kP_\Interp/(\partial \mu_B)^k$ ($k=0,1,2$) 
at the boundaries is represented as $M\vec{a}$, 
where $\vec{a}=(a_n)$ and $M$ is a matrix of $\mu_{B}^L,\mu_{B}^U$. 
Therefore, the $J$ derivative of $a_n$ is calculated as 
\begin{align}
\pdv{a_n}{J}=-M^{-1}\pdv{M}{J}M^{-1}\vec{P}+M^{-1}\pdv{\vec{P}}{J}\,. 
\end{align}
Since $\vec{P}$ is evaluated at the boundaries, 
$\partial\vec{P}/\partial J$ is calculated as, 
for example,
\begin{align}
\pdv{J}\qty(P\big|_{\mu_{B}^{L}})
&
=\pdv{\mu_{B}^{L}}{J}\pdv{P}{\mu_B}\bigg|_{\mu_{B}^{L}} +\pdv{P}{J}\bigg|_{\mu_{B}^{L}}\,,\\
\label{eq-dndJ}
\pdv{J}\qty(\pdv{P}{\mu_B}\bigg|_{\mu_{B}^{L}})
&=\pdv{\mu_{B}^{L}}{J}\pdv[2]{P}{\mu_B}\bigg|_{\mu_{B}^{L}}
+\pdv[2]{P}{J}{\mu_B}\bigg|_{\mu_{B}^{L}}\,,\\
\pdv{J}\qty(\pdv[2]{P}{\mu_B}\bigg|_{\mu_{B}^{L}})
&=\pdv{\mu_{B}^{L}}{J}\pdv[3]{P}{\mu_B}\bigg|_{\mu_{B}^{L}}
+\frac{\partial^3P}{\partial J\partial\mu_B^2}\bigg|_{\mu_{B}^{L}}\,. 
\end{align}
Since the density at the boundaries are fixed for any $J$, 
Eq.(\ref{eq-dndJ}) equals zero and 
\begin{align}\label{eq-dmudJ}
\pdv{\mu_{BL}}{J}=-\pdv[2]{P}{J}{\mu_B}\bigg|_{\mu_{B}^{L}} \bigg/\pdv[2]{P}{\mu_B}\bigg|_{\mu_{B}^{L}}\,. 
\end{align}
Moreover, $\partial^{k+1}P/\partial J (\partial \mu_B)^k =-\partial^k\phi/(\partial \mu_B)^k$ is 
determined from the derivation of the gap equation, for example, 
\begin{align}
0=\pdv{J}\qty(\pdv{\Omega}{\phi}\bigg|_{\phi_\ast})
=\pdv{\phi_\ast}{J}\pdv[2]{\Omega}{\phi}\bigg|_{\phi_\ast}+\pdv[2]{\Omega}{J}{\phi}\bigg|_{\phi_\ast}
\end{align}
where $\phi_\ast$ is the solution of the gap equation.

\bibliography{ref_MKH.bib}

\end{document}